\documentclass[10pt,a4paper]{article}

\usepackage[a4paper,margin=2.2cm]{geometry}
\usepackage{authblk}
\usepackage{microtype}
\usepackage[numbers,square,sort&compress]{natbib}
\usepackage{lineno}
\usepackage{lmodern}
\usepackage[utf8]{inputenc}
\usepackage[T1]{fontenc}
\usepackage{amsmath,amssymb,bm}
\usepackage{graphicx}
\usepackage{subcaption}
\usepackage{booktabs}
\usepackage{siunitx}
\usepackage{physics}
\usepackage{hyperref}
\usepackage{enumitem}
\usepackage{algorithm}
\usepackage{algpseudocode}
\usepackage{xcolor}

\hypersetup{colorlinks=true,allcolors=blue}
\title{Quantum utility routing in continuous-variable QKD networks with trusted and untrusted relays}

\author[1,*]{Farsad Ahmad}
\author[1,+]{Aeysha Khalique}

\affil[1]{Department of Physics, SBA School of Science and Engineering, Lahore University of Management Sciences, Lahore, Pakistan}
\affil[*]{farsad.ahmad@lums.edu.pk}
\affil[+]{aeysha.khalique@lums.edu.pk}

\begin{document}
	\flushbottom
	\maketitle
	
	\begin{abstract}
		Continuous-variable quantum key distribution (CV-QKD) networks require routing strategies that account simultaneously for secret key generation, relay trust and finite network resources. We introduce Quantum Utility Routing (QUR), a weighted routing framework for networks combining trusted point-to-point links with untrusted CV Bell measurement relays. Transition level composable finite size secret key rates (SKRs) are incorporated directly into the route search, allowing QUR to balance bottleneck SKR against route geometry and resource use. We consider graph optimized QUR (QUR$_{\mathrm{go}}$) and a fixed weight implementation (QUR$_{\mathrm{fw}}$) that uses machine learning to avoid repeated graph specific weight optimization, and benchmark both against A* and Max--Min routing. At the default network configuration, QUR$_{\mathrm{go}}$ and QUR$_{\mathrm{fw}}$ improve the mean path SKR relative to A* while requiring substantially fewer network resources than Max--Min. The framework also maintains secure connectivity over a broad range of trusted node availability by exploiting untrusted infrastructure as measurement relays. These results demonstrate that routing in CV-QKD networks can be treated as a configurable trade off between cryptographic performance and network resource consumption.
	\end{abstract}

	\section*{Introduction}
	
	Quantum key distribution (QKD) enables distant users to establish secret keys whose security is derived from the laws of quantum mechanics rather than the computational difficulty of a mathematical problem \cite{qkd1}. Continuous-variable QKD (CV-QKD) is particularly attractive for existing communication networks because it can encode information in the quadratures of coherent or squeezed light and recover it using coherent detection \cite{grosshans2003,cv1,cv2}. CV-QKD utilizes standard optical fiber, well established telecommunication components, high bandwidth coherent receivers and increasingly integrated photonic hardware \cite{cv3,cv4,cv5}. Practical implementations and switched CV-QKD network demonstrations further indicate that CV-QKD is progressing from isolated laboratory links towards a deployable communication infrastructure \cite{jain2022,brunner2023}. The QKD infrastructure of a quantum network can be built from discrete variable (DV) systems, CV systems, or by combining the two. Metropolitan QKD was demonstrated for both DV \cite{elliott2005,sasaki2011} and CV \cite{peev2009} technologies.
	
	
	Most QKD protocols are developed as point-to-point (P2P) systems, whereas practical networks must connect distant users over distances where direct links may become inefficient due to loss, particularly under finite sized security. Hence, extending QKD beyond individual links requires switching, relaying and route selection \cite{pirandola2019,network2,transmissivity1}. This introduces a network-level optimization problem in which the network must determine an optimal way to establish a secure key route. This problem cannot be reduced to physical distance alone, since key generation, key availability, relay placement, resource utilization and trusted--untrusted node constraints can all influence the route \cite{network1,network2,network4,harney}. In CV-QKD, this dependence is further strengthened because the secret key rate (SKR) is determined by channel loss, excess noise, state preparation, detection and reconciliation efficiency, and security assumptions with finite sized effects \cite{cv3,cv4,leverrier2015,jain2022}. In a multi-hop network, the end-to-end key rate is limited by the weakest transition along the route \cite{pirandola2019,harney}. Multipath routing offers a trade off, they can increase the available end-to-end key by combining them or reduce dependence on a single path \cite{network5,multipath1,pirandola2019}, but each additional route also consumes links and nodes that may be required elsewhere in the network. Practical routing must therefore balance key generation performance against network resource use \cite{halder2024}.

	A further complication arises from node trust. Trusted relay networks provide a practical way of extending QKD beyond a single quantum link, but trusted key forwarding requires the intermediate nodes on a route to remain within the trust boundary \cite{network4}. As a network grows and spans infrastructure controlled by different organizations, trusted node assumption can become restrictive. Measurement device independent QKD provides a complementary approach in which the intermediate measurement stations are not trusted \cite{dvmdiqkd1}. In continuous-variable measurement-device-independent QKD (CV-MDI-QKD), remote quantum states can be correlated through a publicly announced CV Bell measurement while the measurement station is assigned to the untrusted party \cite{cvmdiqkd1,cvmdiqkd2}. Composable finite size security has subsequently been developed for CV-MDI protocols and for more general Gaussian quantum networks containing untrusted relays \cite{lupo2018,pirandolanetwork}.
	
	These developments motivate the hybrid network considered here. Trusted nodes establish direct P2P links and untrusted nodes may participate in public CV Bell measurements to connect two trusted nodes. The quality of a transition is determined by total loss and noise, number of relays and their placement, state parameters such as squeezing, and finite sized security penalties \cite{pirandolanetwork}. Route selection over such a topology is inherently multi-objective: a useful route should preserve a high bottleneck SKR while limiting distance, hop count, relay use and unnecessary detour, even though these objectives can conflict. Research in CV-QKD provides realistic models incorporating practical implementation effects \cite{cv3,leverrier2015,jain2022}.
	
	CV QKD networking studies have considered dynamic SKR based routing and shortest path selection with optical resource allocation \cite{transmissivity1, ding2024}. Recent studies have incorporated composable end-to-end security into routing analyses \cite{harney2}. Here, we investigate a routing policy that combines transition level composable finite sized key rates with explicit penalties for distance, relay use, transition count and route geometry, and evaluate the resulting key-rate--resource trade off in mixed trusted--untrusted networks.
	
	To investigate these questions, we introduce \emph{Quantum Utility Routing} (QUR), a weighted routing framework for CV-QKD networks. Each transition is assigned a composable finite size SKR under collective Gaussian attacks, and complete end-to-end routes are formed through trusted node key forwarding, with their rate determined by the weakest transition. QUR combines six routing factors and the resulting utility trades cryptographic performance against the network cost of establishing the route.
	
	As a weighted search, QUR benefits from weight optimization, but repeated optimization is impractical in dynamic networks. We therefore consider graph-optimized QUR (QUR$_{\mathrm{go}}$), where weights are optimized for each graph, and fixed-weight QUR (QUR$_{\mathrm{fw}}$), a learned score model selects a common weight vector for each network configuration. Similar machine learning approaches have been used in QKD to reduce repeated parameter optimization \cite{ml1,ml2}. QUR$_{\mathrm{fw}}$ therefore measures how much of the QUR$_{\mathrm{go}}$ trade off can be retained without repeating the weight optimization on every graph. We compare QUR with two complementary benchmarks. A*, which provides a geometric shortest path reference \cite{astar}, and Max--Min which selects the route with the largest bottleneck SKR \cite{maxmin}.
	
	The principal contribution of this work is an integration of well established physical-layer CV-QKD security with network-layer route selection. The framework embeds composable rates directly into routing, accommodates both trusted P2P operation and untrusted Bell measurement relays, and balances SKR against network resource use under our desired criteria. Hence, it enables a direct comparison between short routes, SKR-maximizing routes and routes that seek a controlled compromise between key rate performance and network efficiency in CV-QKD networks.
	
	\begin{figure}[htbp]
		\centering
		\begin{subfigure}[b]{0.48\textwidth}
			\centering
			\includegraphics[width=0.75\textwidth]{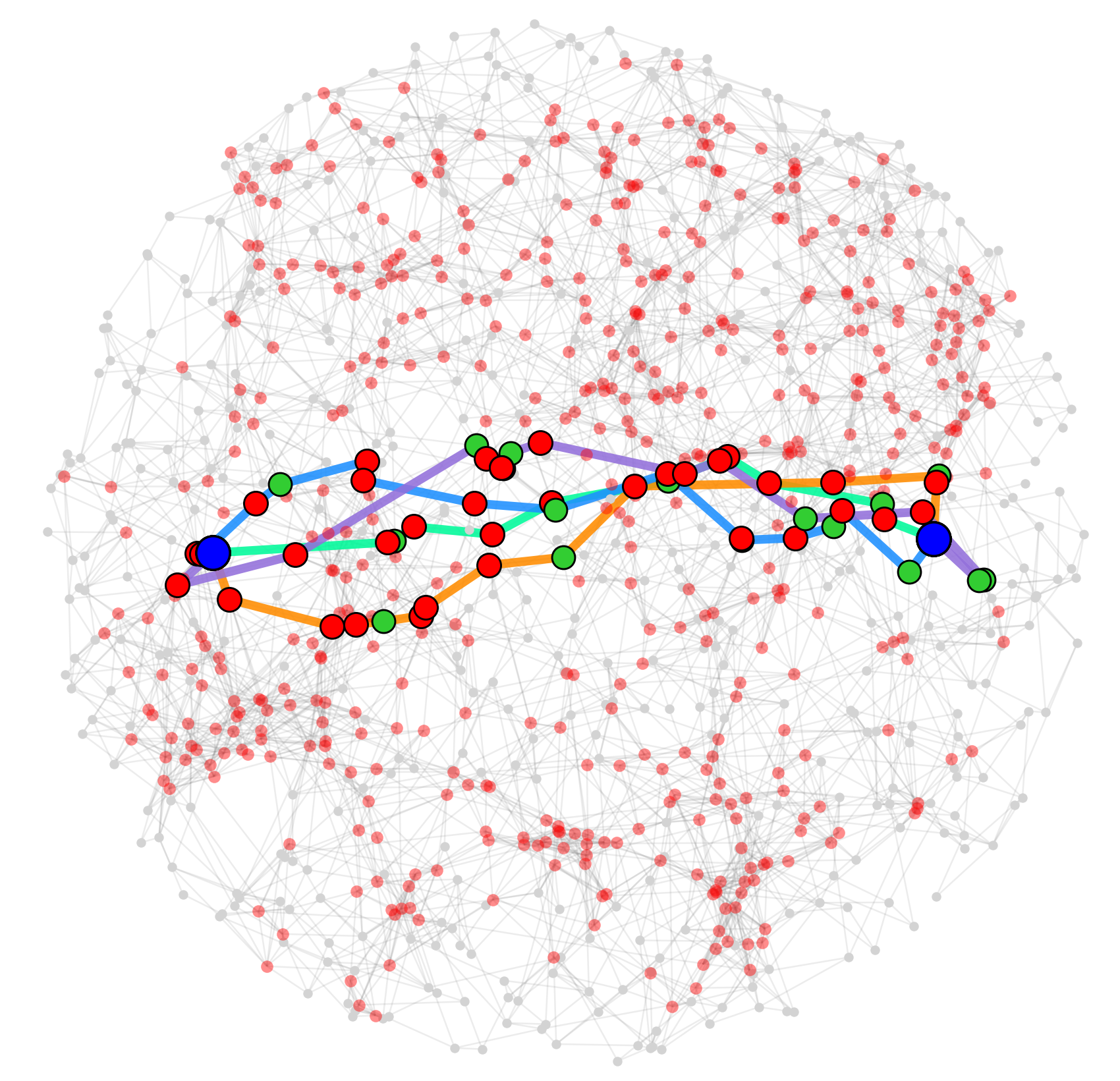}
			\caption{Example Waxman network. Blue nodes mark the source--destination pair, red nodes are untrusted, and green nodes are trusted. The colored curves show the routes returned after successive path extraction.}
			\label{fig:waxman_network}
		\end{subfigure}
		\hfill
		\begin{subfigure}[b]{0.48\textwidth}
			\centering
			\includegraphics[width=0.75\textwidth]{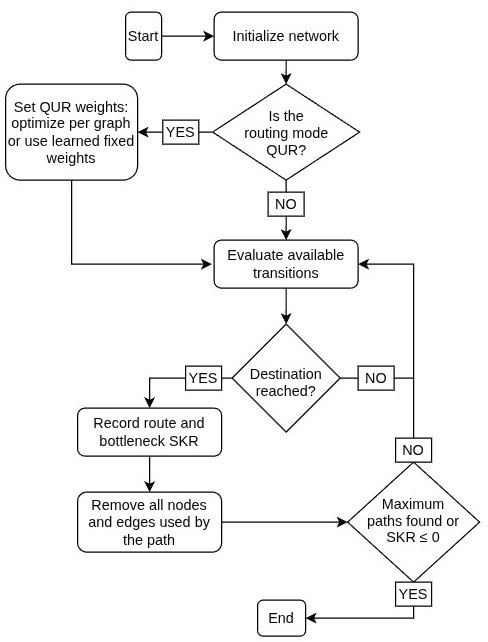}
			\caption{Routing workflow for all algorithms. QUR$_\text{go}$ scores random weight vectors, while QUR$_\text{fw}$ uses the ensemble learned weight vector. A*, Max--Min and both QUR modes then find a route, record its bottleneck SKR, remove nodes and edges used, and continue until the path limit is reached or no positive rate route remains.}
			\label{fig:routing_flowchart}
		\end{subfigure}
		\caption{Simulated network and routing procedure. (a) Example Waxman network at degree threshold $\mathcal{D}=8$, when almost $53$\% of the nodes are untrusted. (b) Workflow used by all routing methods.}
		\label{fig:network_and_flowchart}
	\end{figure}
	
	\section*{Results}
	
	\subsection*{Network setup}
	
	Before stating the results, we briefly describe the setup. We simulate metropolitan scale CV quantum networks using a Waxman random graph \cite{harney}. Nodes are distributed uniformly over a disk and connected according to the Waxman connection probability. Each node represents a network user equipped with two mode squeezed vacuum (TMSV) sources and the required measurement capabilities. Edges represent optical channels through which neighboring users establish entangled links. Consecutive links can be connected by CV Bell measurements at intermediate nodes, with the outcomes publicly announced, thereby producing an effective entanglement-swapped link between the endpoint users. Each graph contains both trusted and untrusted nodes, with trust status determined by node degree: nodes whose degree exceeds a threshold $\mathcal{D}$ are classified as untrusted (see Figure~\ref{fig:waxman_network}).

	The default setting contains 1000 nodes distributed within a disk of radius \SI{50}{km}, a degree threshold $\mathcal{D}=10$, and a maximum of five node disjoint routes per graph. The source and destination are trusted nodes selected near $(-0.7R,0)$ and $(0.7R,0)$, respectively. Each point in a parameter scan is averaged over 50 independently generated graphs.
	
	When the network radius is varied, the link distance is rescaled while maintaining the characteristic distance $r_0=0.07R$. This preserves the graph adjacency structure and only increases the link lengths. If no positive rate route is found, we record zero paths and zero SKR, and include that result in the average. The general workflow of all routing algorithms used in this work is shown in Figure~\ref{fig:routing_flowchart}.
	
	\subsection*{Quantum Utility Routing}
	
	QUR operates on a network whose topology, node positions, link lengths,
	and trust status are known. During the route search, P2P and relay assisted
	transitions between trusted nodes are identified and assigned a transition
	SKR $K_{\ell}$. Transitions with non-positive SKR are excluded.
	
	For a transition $\ell$, QUR assigns the cost
	\begin{equation}
		c_{\ell}^{\mathrm{QUR}}
		=
		q_{\ell}
		+w_d d_{\ell}^{\mathrm{norm}}
		+w_R r_{\ell}
		+w_H h_{\ell}
		+w_A a_{\ell}
		+w_D\delta_{\ell},
		\label{eq:qur_transition_cost}
	\end{equation}
	where
	\[
	q_{\ell}=\max(0,-\log_{10}K_{\ell}).
	\]
	
	Here, $K_{\ell}$ is the SKR of transition $\ell$ evaluated using the security model in Supplementary Note~1 and the parameters in Methods. The six free weights $\bm{w}=(w_d,w_R,w_H,w_A,w_D,w_T)$ determine the relative importance of the cost factors. Their values control the trade off between SKR performance and network resource efficiency. The cost
	factors $d_\ell^{\mathrm{norm}}, r_\ell, h_\ell, a_\ell$ and $\delta_\ell$ describe distance, relay use, hop count, movement away from the
	target, and detour respectively. We explain the mathematics of these coefficients in Methods.
	
	The complete route search procedure is given in Algorithm~\ref{alg:qur}.
    	
	\begin{algorithm}[H]
		\caption{Quantum Utility Routing (QUR)}
		\label{alg:qur}
		\begin{algorithmic}[1]
			
			\Statex \textbf{Input:}
			Network $\mathcal{G}$, source $S$, target $T$, node positions and
			trust states, maximum number of paths $M$, and QUR weight
			vector $\bm{w}=(w_d,w_R,w_H,w_A,w_D,w_T)$.
			
			\Statex \textbf{Output:}
			Up to $M$ routes between $S$ and $T$ and their corresponding
			bottleneck SKRs.
			
			\State \textbf{Step 1: Initialize the search.}
			Begin at the source $S$ with zero accumulated QUR cost and assign
			infinite cost to all other trusted nodes.
			
			\State \textbf{Step 2: Explore transitions from the current node.}
			Identify all P2P and relay assisted transitions from the current
			trusted node to other trusted nodes. For each transition $\ell$,
			determine its SKR $K_\ell$ and calculate its QUR cost
			$c_\ell^{\mathrm{QUR}}$ using Eq.~\ref{eq:qur_transition_cost}.
			
			\State \textbf{Step 3: Update and rank the reachable nodes.}
			For each transition $\ell=(u,v)$, calculate the accumulated cost
			\[
			C_{\mathrm{QUR}}(v)
			=
			\sum_{\ell \in \mathcal R_{S \to u}} c^{\mathrm{QUR}}_\ell
			+
			c_\ell^{\mathrm{QUR}}(v),
			\]
			and rank the reachable nodes using
			\[
			f_{\mathrm{QUR}}(v)
			=
			C_{\mathrm{QUR}}(v)
			+
			w_T\frac{\|\bm{x}_v-\bm{x}_T\|_2}{R}.
			\]
			If more than one route reaches the same node, keep the route with
			the lower accumulated cost. If the costs are equal, keep the route
			with the larger bottleneck SKR.
			
			\State \textbf{Step 4: Explore the next node.}
			Select the node with the lowest $f_{\mathrm{QUR}}$, explore all of
			its outgoing transitions as in Step 2, and rank the reachable nodes
			again. Repeat Steps 2--4 until the target $T$ is reached.
			The SKR of the selected route is
			\[
			K_{\mathcal R}
			=
			\min_{\ell\in\mathcal R}K_\ell .
			\]
			
			\State \textbf{Step 5: Search for additional routes.}
			Record the selected route, exclude its used physical edges and
			internal nodes, and repeat Steps 1--4 until $M$ routes are
			found or no further route is available.
			
		\end{algorithmic}
\end{algorithm}

    $\ell=(u,v)$ denotes a transition from the
	currently explored trusted node $u$ to a reachable trusted node $v$.
	The vector $\bm{x}_u$ gives the spatial position of node $u$, while
	$\bm{x}_T$ gives the target position. $\|\cdot\|_2$ denotes the
	Euclidean norm and $R$ denotes the network radius.

	For comparison, the search priorities of QUR, A* and Max--Min for a partial route are
	\begin{align}
		f_{\mathrm{QUR}}(v)
		&=
		C_{\mathrm{QUR}}(v)
		+
		w_T\frac{\|\bm{x}_v-\bm{x}_T\|_2}{R},
		\nonumber\\
		f_{A^*}(v)
		&=
		\sum_{\ell \in \mathcal R_{S \to v}}d_\ell
		+
		\|\bm{x}_v-\bm{x}_T\|_2,
		\nonumber\\
		f_{\mathrm{MM}}(v)
		&=
		\max_{\mathcal R\in\mathcal R_{S\rightarrow v}}
		\min_{\ell\in\mathcal R}K_{\ell}.
		\label{eq:routing_priorities}
	\end{align}
	Thus, A* primarily favors geometric progress, Max--Min favors the
	strongest bottleneck SKR, while QUR combines transition quality,
	route geometry, and network resource use. The weight vector
	$\bm{w}=(w_d,w_R,w_H,w_A,w_D,w_T)$ is specific to QUR. Neither A* nor
	Max--Min uses an optimized weight vector.
	
	Two QUR operating modes are considered. In graph optimized QUR
	(QUR$_{\mathrm{go}}$), the six weights are optimized independently
	for each generated network using Monte Carlo sampling. In fixed weight
	QUR (QUR$_{\mathrm{fw}}$), a histogram based gradient boosting model
	is trained using graph characteristics and the performance of randomly
	sampled weight vectors, from which a common weight vector is obtained
	for each network configuration. Hence, the distinction between QUR$_{\mathrm{go}}$ and QUR$_{\mathrm{fw}}$ lies only in how the weight vector is selected. For both modes, the weight vector optimization follows Algorithm~\ref{alg:qur_weights}.

	\begin{algorithm}[H]
		\caption{Selection of QUR weight vectors}
		\label{alg:qur_weights}
		\begin{algorithmic}[1]
			
			\Statex \textbf{Input:}
			Network ensemble and QUR weight space $\mathcal W$.
			\Statex \textbf{Output:}
			Graph-specific weights for QUR$_{\mathrm{go}}$ and one common
			weight vector for QUR$_{\mathrm{fw}}$ for the network configuration.
			
			\State \textbf{Step 1: Evaluate sampled weights.}
			For each graph, sample $8{,}000$ distinct weight vectors and
			evaluate each vector using Algorithm~\ref{alg:qur}. Assign each
			vector the score $S(\bm w)$ defined in
			Eq.~\ref{eq:optimizer_score}.
			
			\State \textbf{Step 2: Select QUR$_{\mathrm{go}}$.}
			For each graph independently, select the weight vector with the
			highest score.
			
			\State \textbf{Step 3: Learn the fixed weight model.}
			Train the histogram based gradient boosting model using the graph
			properties, tested weight vectors, and their transformed scores.
			
			\State \textbf{Step 4: Select QUR$_{\mathrm{fw}}$.}
			Form a common weight pool of the highest scoring vectors of each
			graph with additional random vectors. Evaluate their
			predicted performance and select the
			vector with the highest mean relative performance.
			
		\end{algorithmic}
\end{algorithm}

    The weight space $\mathcal W$ used in this work can be found in Methods.

	\begin{figure}[htbp]
		\centering
		\includegraphics[width=\linewidth]{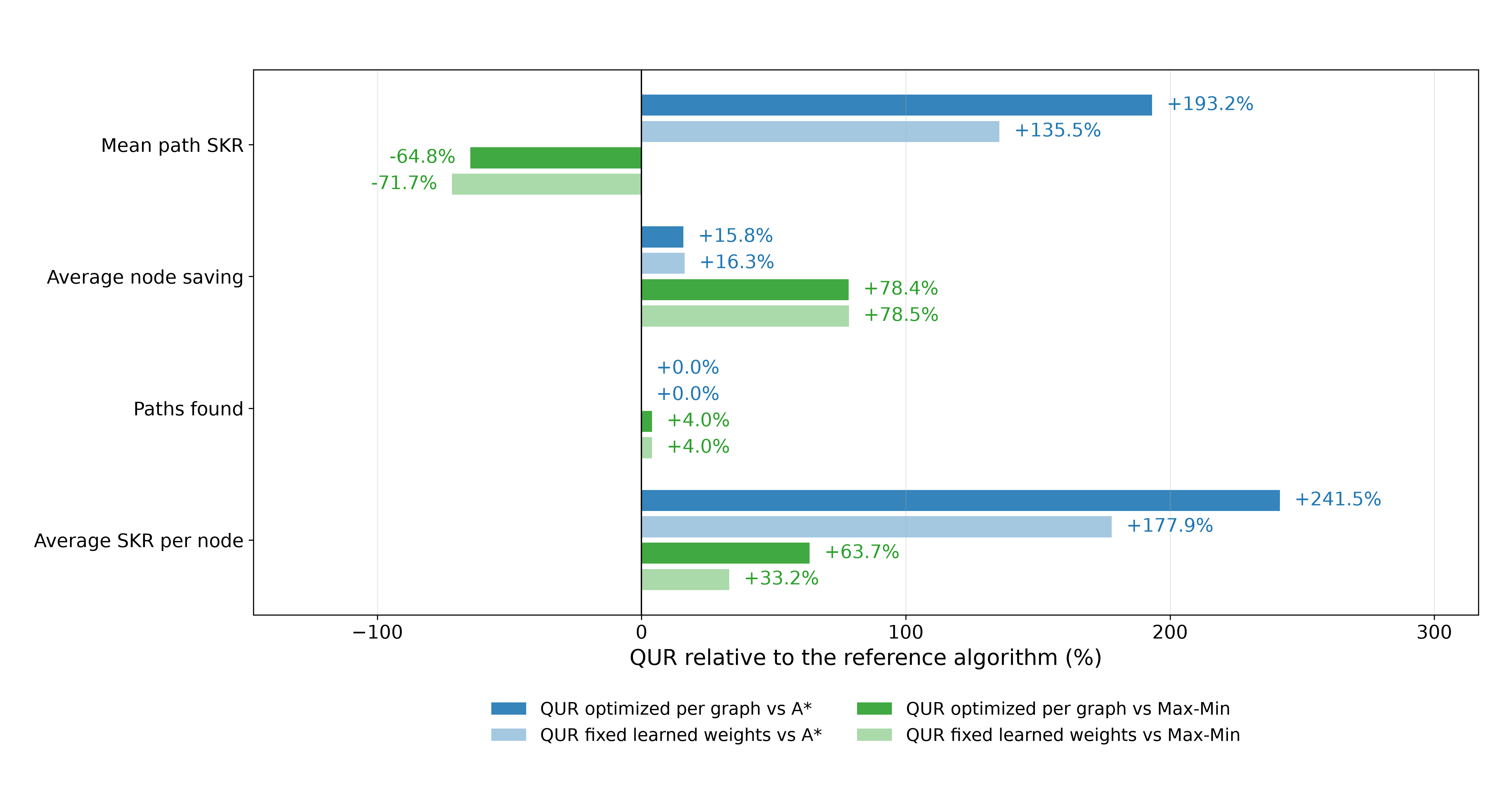}
		\caption{Overall comparison at the default network setting. Bars report QUR$_\text{go}$ and QUR$_\text{fw}$ relative to A* or Max--Min. Average node saving is the mean of internal and unique node saving. Average SKR per node is the mean of relative changes in the two efficiency measures defined in Eqs.~\ref{eq:internal_efficiency} and \ref{eq:unique_efficiency}. Positive values favor the corresponding QUR mode.}
		\label{fig:overall_qur_comparison}
	\end{figure}
	
	\subsection*{Routing performance and resource efficiency}
	
	Figure~\ref{fig:overall_qur_comparison} reveals three distinct routing regimes. A* preferentially selects geometrically short routes, but these routes can contain weak quantum transitions and therefore provide relatively low bottleneck SKR. Max--Min takes the opposite approach, it maximizes the weakest transition along the route, but does this at the cost of substantially larger node use. QUR occupies the intermediate operating regime, improving the bottleneck SKR of the short routes favored by A* while avoiding the large resource requirement associated with Max--Min.
	
\begin{figure}[htbp]
		\centering
		\includegraphics[width=0.9\linewidth]{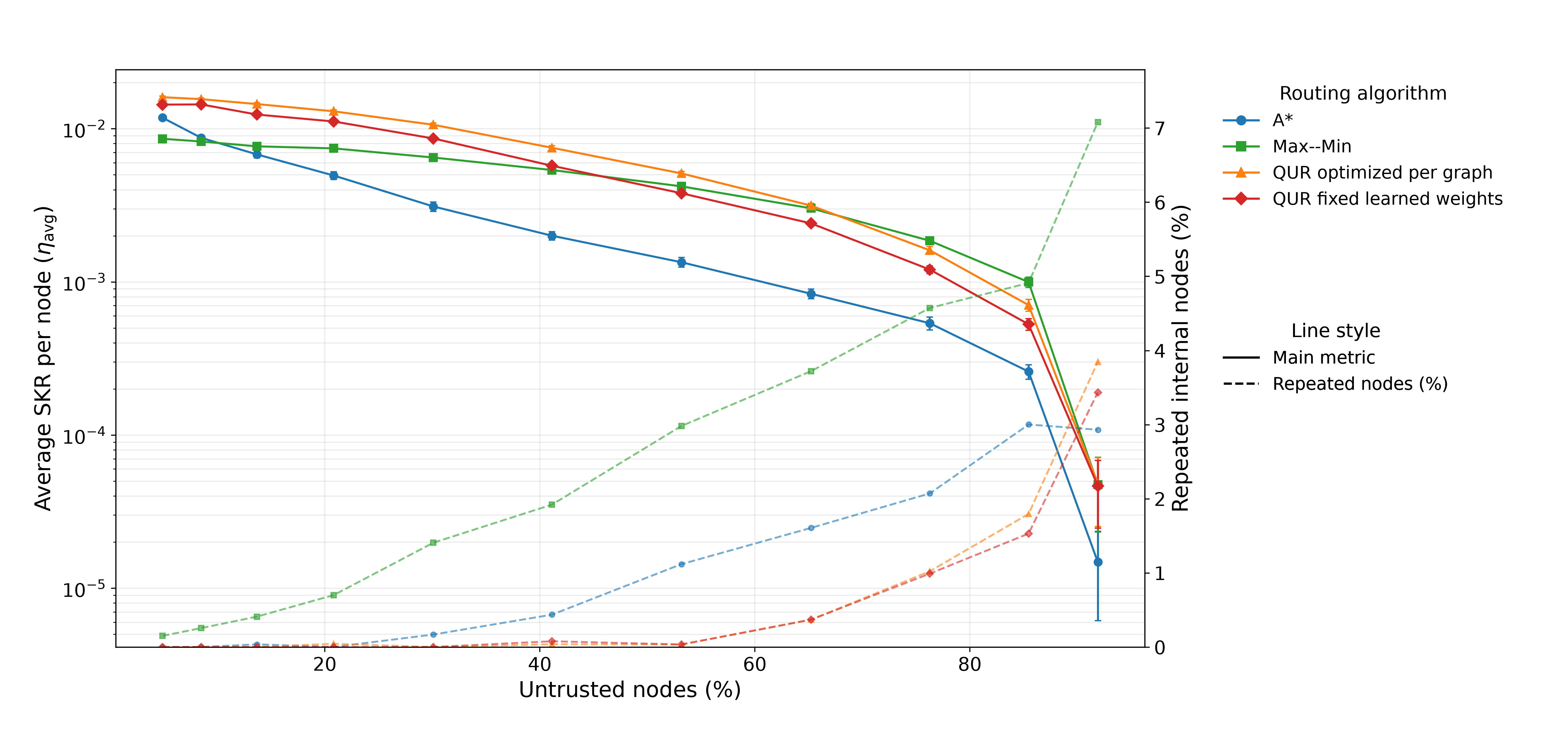}
		\caption{Average SKR efficiency and node reuse as functions of the untrusted node percentage. Solid curves show $\eta_{\mathrm{avg}}$ from Eq.~\ref{eq:average_efficiency}. Dashed curves show repeated nodes as a percentage of all nodes.}
		\label{fig:compromised_nodal_efficiency}
	\end{figure}

	This distinction becomes clearer when we look at SKR efficiency from Figure~\ref{fig:compromised_nodal_efficiency}. Both QUR modes outperform the reference algorithms in the SKR-per-node comparison, showing that their advantage is not simply associated with short routes or high rate routes individually. Rather, QUR improves the SKR obtained from the intermediate node resources available to a connection.
	
	At the default configuration, QUR$_{\mathrm{go}}$ increases the mean path SKR by 193.2\% relative to A*, while using 78.4\% fewer nodes than Max--Min. QUR$_{\mathrm{fw}}$ shows the same qualitative behavior.
	
	Importantly, this resource advantage is not obtained by reducing multipath availability. The mean number of accepted paths remains similar across the routing methods. QUR therefore achieves its improved resource efficiency primarily by changing which routes are selected, rather than by accepting fewer routes.
	
	The fixed weight result is important because removing graph specific optimization changes the resource use behavior of QUR marginally. Although QUR$_{\mathrm{fw}}$ sacrifices some SKR relative to QUR$_{\mathrm{go}}$, it retains almost the same node saving and multipath behavior. This indicates that the useful QUR operating regime is not produced only by highly specialized weights for individual graphs. Instead, a common weighting vector appears to capture much of the routing behavior across the network ensemble.
	
	\subsection*{Routing through highly untrusted networks}
	
	A node $v$ is labeled untrusted when its degree satisfies $\deg(v)>\mathcal{D}$. Decreasing $\mathcal{D}$ reduces the trusted node population and provides a controlled way of testing routing when much of the network cannot participate in trusted key forwarding.
	
	We can observe from Figure~\ref{fig:compromised_nodal_efficiency} and Figure~\ref{fig:multipath_reuse_scaling} that increasing the untrusted node fraction does not immediately destroy secure
	connectivity. Because untrusted nodes can still participate as public Bell
	measurement relays, they remain useful for establishing transitions between
	trusted nodes rather than being removed entirely from the network. From our simulations, we observe that at
	$\mathcal{D}=6$, where 76.3\% of the nodes are untrusted,
	QUR$_{\mathrm{go}}$ still achieves 100.0\% routing success and
	finds approximately 3.92 routes per graph. Thus, a trusted minority
	can continue to support end-to-end key distribution when the surrounding
	untrusted infrastructure remains available as untrusted relays.
	
	A different regime appears at the most severe trust
	restriction. At $\mathcal{D}=4$, we recorded that approximately 91.9\% of the
	network is untrusted and routing success falls sharply. At this point the
	limiting factor is no longer only the SKR of relay assisted
	transitions but also that the remaining trusted nodes become too sparse to form a positive SKR connection between the endpoints.
	
	Hence the trust scan shows a transition from a rate-limited regime to a trusted-connectivity-limited regime in which too few
	trusted nodes remain to support end-to-end routing. Allowing untrusted Bell
	measurement relays substantially extends the former regime, but cannot
	compensate indefinitely.
	
	\subsection*{Multipath capability}
	
	Max--Min can leave fewer residual routes because a high bottleneck route may consume many nodes before the next search begins. On a graph-by-graph basis, both modes of QUR match the A* path count in 100.0\% of graphs and the Max--Min path count in 84.0\%.
	
	Multiple node disjoint routes can be used in two ways. Their keys may be used independently to increase the end-to-end SKR, or path keys may be combined at the endpoints,
	\begin{equation}
		k_{\mathrm{final}}
		=
		k_1\oplus k_2\oplus\cdots\oplus k_m,
	\end{equation}
	where $k$ is the final key string. If the path keys are independently generated and at least one remains secret conditioned on the other path keys and
	Eve's information, the XOR key remains secret. This use requires authenticated communication, equal key lengths and trusted endpoint processing \cite{harney,network5}. In the default setting, both QUR modes match the A* path count on all tested graphs and match the Max--Min path count on most graphs. So the node saving obtained by QUR is not simply a result of accepting fewer paths.
	
	\begin{figure}[htbp]
		\centering
		\includegraphics[width=0.4\linewidth]{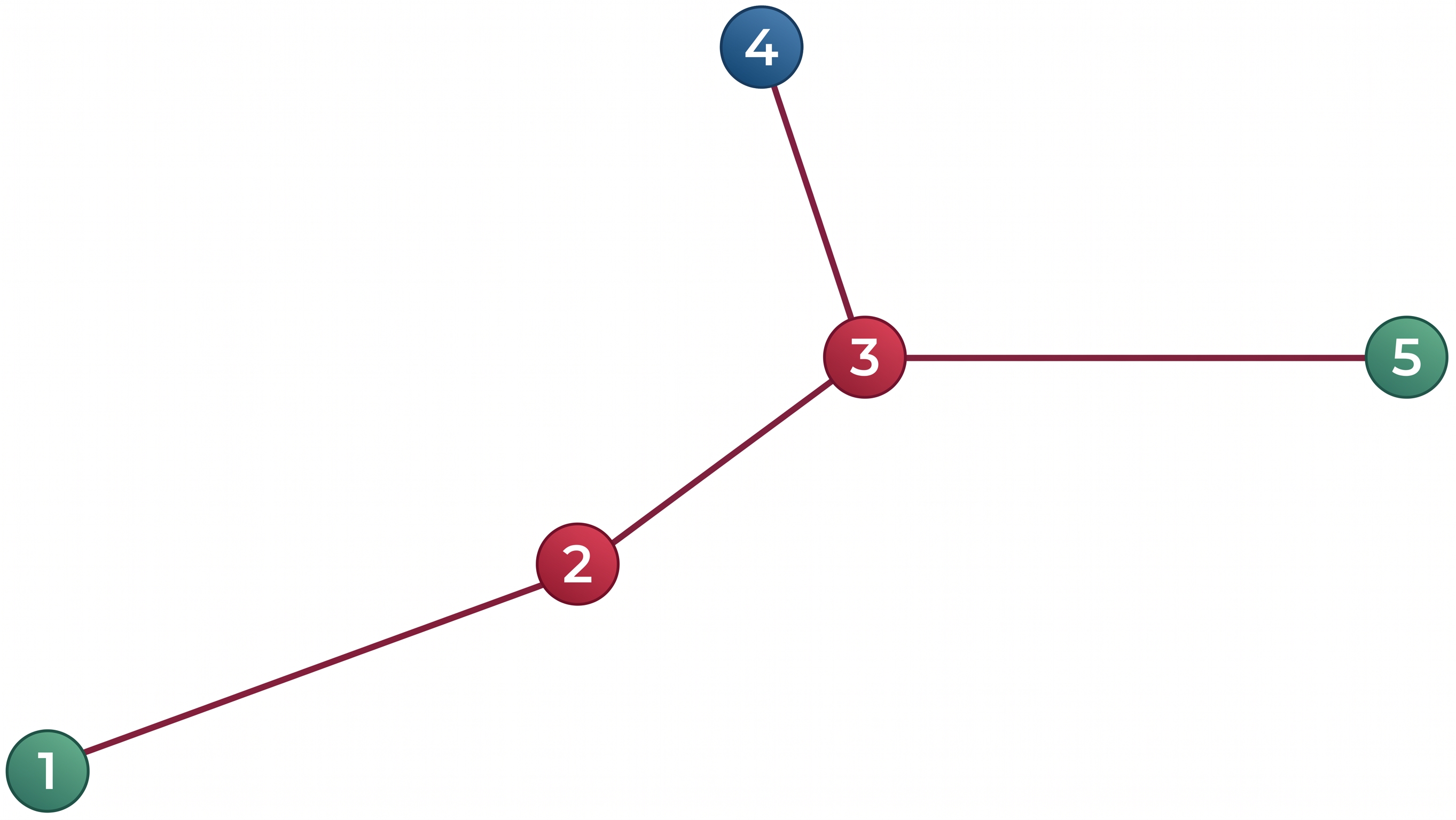}
		\caption{Example of within-route node reuse. Green nodes are trusted and red nodes are untrusted. The path constructed from the route is $1\!\rightarrow\!2\!\rightarrow\!3\!\rightarrow\!4\!\rightarrow\!3\!\rightarrow\!5$, in which node 3 appears in two separate transitions i.e. $1\!\rightarrow\!2\!\rightarrow\!3\rightarrow\!4$ and $4\!\rightarrow\!3\!\rightarrow\!5$.}
		\label{fig:nodereuse}
	\end{figure}
	
	\subsection*{Node reuse and SKR efficiency}	
	
	The SKR of a route does not by itself describe how efficiently the network is being used so we evaluate the SKR obtained per node. Since nodes and edges may be reused within the same route, we use two measures of node usage. The first counts every node in the path, including repeated ones, labeled internal nodes while the second counts only the distinct nodes used by all paths labeled unique nodes. For each graph, the internal node efficiency is defined as
	\begin{equation}
		\eta_{\mathrm{int}}
		=
		\begin{cases}
			\overline{K}/\overline{n}, & \overline{n}>0,\\
			0, & \overline{n}=0,
		\end{cases}
		\label{eq:internal_efficiency}
	\end{equation}
	where $\overline{K}$ is the mean bottleneck SKR of the accepted paths and $\overline{n}$ is their mean number of nodes. For the unique nodes, we define
	\begin{equation}
		\eta_{\mathrm{unique}}
		=
		\begin{cases}
			K_{\mathrm{sum}}/N_{\mathrm{unique}}, &
			N_{\mathrm{unique}}>0,\\
			0, & N_{\mathrm{unique}}=0,
		\end{cases}
		\label{eq:unique_efficiency}
	\end{equation}
	where $K_\mathrm{sum}$ is the sum of bottleneck SKR of each path and $N_{\mathrm{unique}}$ is the total number of unique nodes used by all paths. Their mean,
	\begin{equation}
		\eta_{\mathrm{avg}}
		=
		\frac{\eta_{\mathrm{int}}+\eta_{\mathrm{unique}}}{2},
		\label{eq:average_efficiency}
	\end{equation}
	is used as the overall SKR-per-node measure.
	
	We can observe from Figure~\ref{fig:compromised_nodal_efficiency} that generally, both modes of QUR offer a higher key rate per node used when compared to the two reference algorithms. Importantly, this improvement is not obtained simply by returning fewer paths. The mean number of accepted paths is similar across the routing algorithms, and QUR$_{\mathrm{go}}$ matches the A* path count in 100.0\% of the tested graphs. Thus, the higher SKR-per-node efficiency reflects a different choice of routes rather than an artificial reduction in multipath availability.
	
\begin{figure}[htbp]
		\centering
		\includegraphics[width=1.05\linewidth]{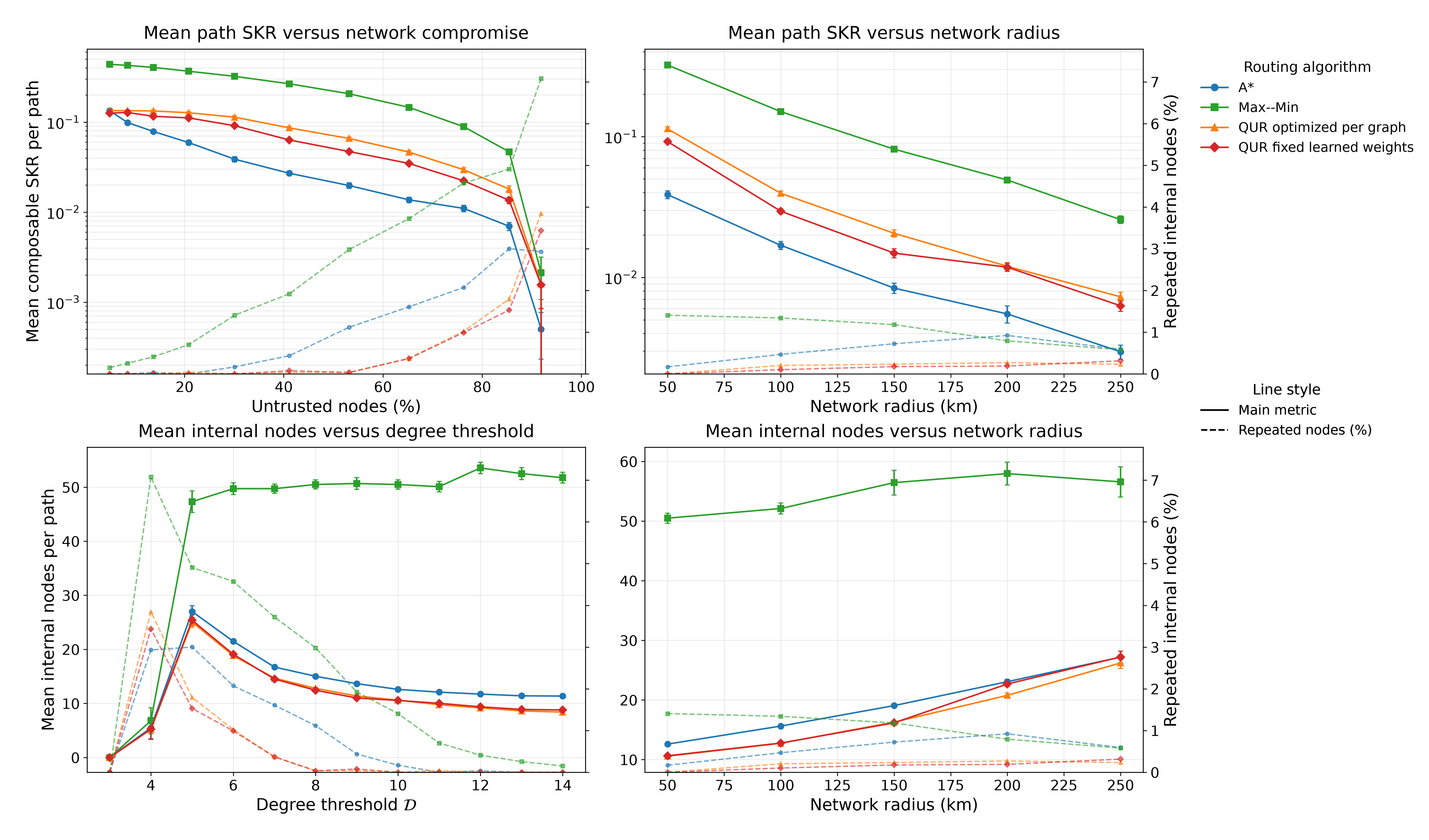}
		\caption{Routing performance and within-route node reuse. The upper panels show the mean composable SKR per path against untrusted node percentage and network radius. The lower panels show the mean number of nodes per path against degree threshold and radius. Dashed curves and the right axes show repeated internal node percentage.}
		\label{fig:multipath_reuse_scaling}
	\end{figure}

	Figure~\ref{fig:compromised_nodal_efficiency} shows how $\eta_{\mathrm{avg}}$ changes as the fraction of untrusted nodes increases, while Figure~\ref{fig:multipath_reuse_scaling} compares the corresponding mean path SKR and node requirements across the $\mathcal D$ and $R$ scans. Both figures also show the percentage of nodes reused isnide each accepted path. As expected, node reuse increases as $\mathcal D$ decreases because fewer trusted intermediate stations remain available for routing. This effect is particularly pronounced for Max--Min, whose high nodal requirement rapidly depletes the set of available trusted nodes when multiple paths are constructed.

	\section*{Discussion}
	
	The main implication of these results is that secure routing in a CV-QKD network should not be treated as either a shortest path or a maximum SKR problem alone. These objectives represent different operating extremes: reducing geometric distance does not guarantee strong quantum transitions, whereas maximizing the bottleneck SKR can consume a large fraction of the available network resources. QUR instead treats routing as a configurable trade off between physical-layer key generation and network-layer resource use. The significance of the framework is not a single universally optimal set of weights, but the ability to tune the routing policy according to the requirements of the network.
	
	This flexibility is important for shared quantum networks. Intermediate stations, optical hardware and available trusted nodes are finite resources, and consuming them for one connection can restrict the routes available to subsequent users or paths. From this perspective, node use is not merely an auxiliary routing statistic but an operational cost. The SKR-per-node metric introduced here provides a complementary measure to the conventional bottleneck SKR by asking how effectively the available infrastructure is converted into secure key capacity.
	
	The behavior of node reuse provides a related indication of network pressure. Repeated use of intermediate nodes can allow a route to remain feasible when trusted connectivity becomes limited, but increasing reuse also indicates that fewer structurally distinct alternatives remain. Node reuse should therefore be interpreted both as a mechanism for maintaining connectivity and as an indicator that the routing space is becoming constrained. This becomes particularly relevant for multipaths, where excessive consumption of useful nodes by an early route can reduce the opportunities available to later routes.
	
	The hybrid trusted--untrusted architecture also has an important network design implication. An untrusted node does not necessarily need to be removed from the communication infrastructure. By allowing such nodes to participate through public CV Bell measurements, they can continue to support connectivity without being entrusted with the final end-to-end key. The ultimate limitation is then determined by the availability and spatial distribution of trusted nodes and the feasibility of relay assisted transitions. Consequently, the placement of trusted stations may be as important as their overall fraction when designing large QKD networks.
	
	The distinction between QUR$_{\mathrm{go}}$ and QUR$_{\mathrm{fw}}$ is primarily operational. Graph specific optimization provides a useful reference for the best QUR
	performance identified within the $8{,}000$ sample search budget. The idea is motivated from the fact that repeating a large weight search whenever network conditions change would be difficult in practice. QUR$_{\mathrm{fw}}$ instead suggests that the expensive optimization stage may be moved offline, leaving only the route search and transition rate updates to be performed during network operation. This is particularly relevant to dynamic CV-QKD networks, where channel transmissivity and therefore the available SKR can vary with time \cite{transmissivity1}.
	
	The score function should be viewed as a network policy choice rather than a fixed objective. In the present implementation, the optimization score favors reduced node use while discouraging solutions with poor key rate. A network operator could instead emphasize other requirements, such as relay avoidance, higher key throughput, lower latency, multipath availability or resilience. The weighted formulation therefore provides a mechanism for incorporating network-level priorities without modifying the underlying physical-layer security model.
	
	The fixed-weight vector is learned and evaluated on the same ensemble of Waxman network. Present results establish that a common weight vector can represent that ensemble. The methodology described here can extend to unseen topologies. We can also extend the simulations to include other factors such as heterogeneous equipment, time dependent channels and trust assignments based on ownership, location or operational history.
	
	In the present framework, composability is applied at the transition level. Recent work has shown that end-to-end composable security introduces a security--rate trade off in which a total security budget must be distributed across all of the transitions forming the route \cite{harney2}. This creates a coupling between routing and security allocation: the number and type of transitions are not known until a route has been constructed, while their achievable SKRs depend on the security budget assigned to all of them. This coupling is avoided by assigning a fixed and relatively strict security parameter to each transition. For a route $\mathcal R$ containing $L_{\mathcal R}$ transitions,
	composable security gives the route-level bound
	\begin{equation}
		\epsilon_{\mathcal R}
		\leq
		\sum_{\ell\in\mathcal R}\epsilon_{\mathrm{tr},\ell}.
	\end{equation}
	For the uniform transition security used here, this reduces to
	$\epsilon_{\mathcal R}\leq
	L_{\mathcal R}\epsilon_{\mathrm{tr}}$. Consequently, the end-to-end security parameter is not fixed a priori, but increases with the number of transitions forming the selected route.
	
	\section*{Methods}
	
	\subsection*{Transition rate evaluation and parameters}

We evaluate direct and relay assisted transitions using the established
Gaussian network security model under collective Gaussian attacks
\cite{pirandolanetwork}. The covariance matrix and composable finite size expressions are given in Supplementary Note~1. A transition contains $L_\ell=1,\ldots,4$ physical links and $L_\ell-1$ untrusted relays.
The noise $\xi$ is output referenced, $\eta_{\mathrm B}$ is the Bell-measurement
efficiency, and the fiber attenuation is $\alpha=0.2~\mathrm{dB/km}$.

	Throughout the work, we use
	\begin{equation}
		\mathcal N=10^9,\quad
		f_{\mathrm{PE}}=0.10,\quad
		\beta=0.98,\quad
		\mathrm{FER}=0.10,\quad
		|\mathcal X|=32,\quad
		\xi=0.01,\quad
		\eta_{\mathrm B}=0.98,
	\end{equation}
	and
	\begin{equation}
		\epsilon_{\mathrm{PE}}
		=
		\epsilon_s
		=
		\epsilon_h
		=
		\epsilon_c
		=
		10^{-10}.
	\end{equation}
	For the parameters used here, this gives
	$\epsilon_{\mathrm{tr}}=3.9\times10^{-10}$. For P2P transitions, the SKR is optimized over $r_{\mathrm{dB}}\in[1,30]$ dB.

	Optimizing nested relay transitions is computationally intensive, so we use a discrete hill climb search. We initialize at $r_{\mathrm{dB}}=14$ dB and evaluate the SKR at the neighboring values $r_{\mathrm{dB}}\pm1$ dB. If either neighbor give a higher SKR, the search proceeds in that direction in increments of $1$ dB until no further improvement is obtained or the boundary $1\leq r_{\mathrm{dB}}\leq30$ is reached.
	
	\subsection*{Network model and simulation parameters}

We use an undirected Waxman graph $G=(V,E)$ with $N_V = |V|$ nodes
distributed uniformly within a disk of radius $R$. The sampling
and connection probability formulas are given in Supplementary Note~2
\cite{harney}.

	Node trust is assigned from its degree, with node $v$ treated as untrusted when
	\begin{equation}
		\deg(v)>\mathcal D,
		\label{eq:compromise}
	\end{equation}
	where $\mathcal D$ is the trust threshold.
	
	The default settings are
	\begin{equation}
		N_V=1000,
		\qquad
		R=\SI{50}{km},
		\qquad
		\beta_W=1,
		\qquad
		r_0=0.07R,
		\qquad
		\mathcal D=10,
		\qquad
		M=5,
		\label{eq:default_network_parameters}
	\end{equation}
	where $M$ is the maximum number of routes considered.
	
	To examine the effects of network scale and trust, we independently vary
	\begin{equation}
		R\in{50,100,150,200,250}\ \mathrm{km},
	\end{equation}
	and
	\begin{equation}
		\mathcal D\in{14,13,12,11,10,9,8,7,6,5,4,3},
	\end{equation}
	while keeping all remaining parameters fixed at their default values. Each parameter point is averaged over 50 independently generated graphs per setting.

	\subsection*{Route and multipath construction}
	
	A route $\mathcal R$ is formed as a sequence of transitions between trusted
	nodes. Once a route is found, each transition is expanded into its underlying
	node sequence to obtain the corresponding path $p$, defined as the ordered
	sequence of all nodes used by that route.
	
	For a route $\mathcal R$ consisting of transitions $\ell$, the end-to-end
	SKR is limited by its weakest transition,
	\begin{equation}
		K(\mathcal R)
		=
		\min_{\ell\in\mathcal R}K_{\ell}.
		\label{eq:path_bottleneck}
	\end{equation}
	Rates are expressed in secret bits per transition use. The path $p$ inherits the same end-to-end rate.
	
	Multipath construction is iterative. After a route is accepted, it is expanded
	into its path and every node and edge appearing in that path is removed from the network. The source and destination are retained. The routing search is then repeated until $M$ routes are found or no positive rate route remains. Accepted paths are therefore internally node and edge disjoint from one another.
	
	For $m$ accepted routes $\mathcal R_j$ and their corresponding paths $p_j$, we define the following quantities
	\begin{align}
		K_{\mathrm{sum}}
		&=
		\sum_{j=1}^{m}K(\mathcal R_j),
		\nonumber\\
		\overline K
		&=
		\frac{1}{m}
		K_\mathrm{sum},
		\nonumber\\
		\overline n
		&=
		\frac{1}{m}
		\sum_{j=1}^{m}n(p_j),
		\label{eq:multipath_metrics}
	\end{align}
	where $K_\mathrm{sum}$ is the sum of bottleneck key rates, $\overline K$ is the average SKR per path, and $n(p_j)$ is the number of nodes in the path, including repeated appearances of the same node. This gives us $\overline n$ as the average number of nodes used by those paths. The source and destination are excluded from this count. A simulation in which no route is found is recorded with $\overline K = \overline n = p = 0$.
	
The reference algorithms are geometric A* routing \cite{astar}
and maximum bottleneck Max--Min routing \cite{maxmin}, described in
Supplementary Notes~3 and~4, respectively. Both use the same transition
rates and iterative resource removal procedure defined above.

	\subsection*{Quantum Utility Routing cost function and search priority}
	
	For a transition $\ell=(u,v)$, where $u$ is the current trusted node and
	$v$ is a reachable trusted node, we define
	\begin{align}
		q_{\ell}
		&=
		\max(0,-\log_{10}K_{\ell}),
		\nonumber\\
		r_{\ell}
		&=
		\text{number of untrusted relays in }\ell,
		\nonumber\\
		\delta_{\ell}
		&=
		\frac{1}{R}
		\max\!\left(
		0,
		d_{\ell}
		-
		\|\bm x_u-\bm x_v\|_2
		\right),\nonumber \\
		a_{\ell}
		&=
		\begin{cases}
			0, & r_{\ell}=0, \\[6pt]
			\displaystyle
			\frac{1}{R}
			\sum_{j}
			\max\!\left[
			0,
			\|\bm x_{j+1}-\bm x_T\|_2
			-
			\|\bm x_j-\bm x_T\|_2
			\right],
			& r_{\ell}\geq1,
		\end{cases}
	\end{align}
	where $d_{\ell}$ is the total physical length of the transition and the index $j$ runs over its ordered node sequence. Physically, $q_\ell$ penalizes low SKR, $r_\ell$ counts relay use, $a_\ell$ penalizes backward movement within relay transitions, and $\delta_\ell$ penalizes relay detour from the direct distance between the trusted endpoints. For a direct P2P transition, $r_{\ell}=a_{\ell}=\delta_{\ell}=0$. From this, the QUR cost assigned to transition $\ell$ is given as
	\begin{equation}
		c_{\ell}^{\mathrm{QUR}}
		=
		q_{\ell}
		+w_d\frac{d_{\ell}}{R}
		+w_Rr_{\ell}
		+w_H
		+w_Aa_{\ell}
		+w_D\delta_{\ell}.
		\label{eq:qur_transition_cost_full}
	\end{equation}
	Here, the constant term $w_H$ assigns one hop cost to each transition. For a partial route $\mathcal R_{S\rightarrow u}$, extending it from $u$ to $v$ through
	$\ell=(u,v)$ gives
	\begin{equation}
		C_{\mathrm{QUR}}(v)
		=
		\sum_{\ell\in\mathcal R_{S\rightarrow u}}
		c_{\ell}^{\mathrm{QUR}}
		+
		c_{\ell}^{\mathrm{QUR}}(v).
	\end{equation}
	
	The search priority assigned to the reachable node $v$ is
	\begin{equation}
		f_{\mathrm{QUR}}(v)
		=
		C_{\mathrm{QUR}}(v)
		+
		w_T
		\frac{\|\bm x_v-\bm x_T\|_2}{R}.
		\label{eq:qur_priority}
	\end{equation}
	
	QUR is a heuristic search whose additive cost represents the desired trade off between transition quality and resource use. For each reachable node, the route with the lower accumulated QUR cost is retained. If the costs are equal, the route with the larger bottleneck SKR is preferred. The reachable node with the lowest $f_{\mathrm{QUR}}(v)$ is explored next. The six free QUR weights that control the penalties are
	\begin{equation}
		\bm w
		=
		(w_d,w_R,w_H,w_A,w_D,w_T).
	\end{equation}

	\subsection*{Monte Carlo selection of graph specific QUR weights}
	
	Every weight in $\bm w$ is independently sampled from
	\begin{equation}
		\mathcal W=\{0.00,0.01,\ldots,1.00\}.
		\label{eq:weight_grid}
	\end{equation}
	This creates a space containing
	\begin{equation}
		|\mathcal W|^6=101^6=1{,}061{,}520{,}150{,}601
	\end{equation}
	possible vectors. Checking every combination is hence impractical. For each graph, $8{,}000$ distinct vectors are sampled from this grid and evaluated by the QUR routing model. Every weight vector searches for up to $M$ routes.
	
	We associate a score $S(\bm w)$ with the performance of each sampled wight vector. A vector with no positive rate route receives score $-\infty$. For $0<\overline K<1$ and $\overline n>0$, the base score is
	\begin{equation}
		S_0(\bm w)
		=
		-\frac{1}{\overline n(\bm w)\log_{10}\overline K(\bm w)}.
		\label{eq:base_score}
	\end{equation}
	If a direct source--destination route gives $\overline n=0$, we assign an infinite base score and hence we always prefer it.
	
\begin{figure}[htbp]
		\centering
		\includegraphics[width=0.6\linewidth]{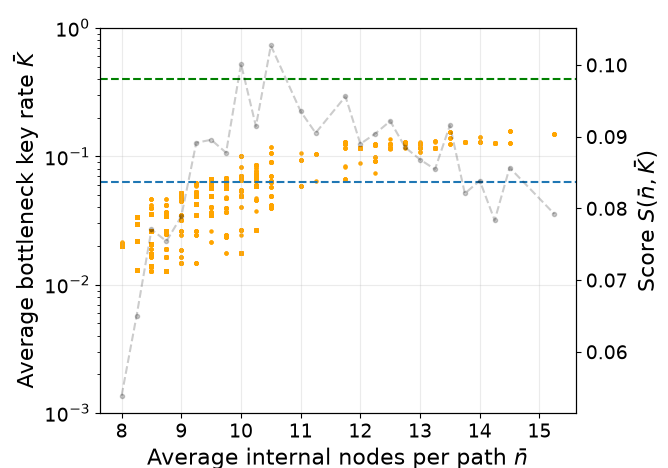}
		\caption{Example Monte Carlo weight search on one graph. Each orange point is one sampled QUR weight vector, shown by its mean node $\bar n$ count and mean bottleneck SKR $\bar K$. They gray dashed line is the score value plotted for the largest $\bar K$ at each $\bar n$. The A* and Max--Min mean rates are shown as the dashed blue and green reference lines respectively. In this simulation, $8,000$ vectors are sampled and evaluated on a graph created at default values. On average, A* used 11.75 nodes and Max--Min used 49.}
		\label{fig:montecarlo}
	\end{figure}

	The A* rate penalty is
	\begin{equation}
		\delta_{A^*}(\bm w)
		=
		\max\!\left[
		0,
		\frac{\overline K_{A^*}-\overline K(\bm w)}
		{\overline K_{A^*}}
		\right],
	\end{equation}
	where $\overline K_{A^*}$ is the mean bottleneck SKR of the routes returned by A*, and $\overline K(\bm w)$ is the corresponding mean bottleneck SKR obtained by QUR using the weight vector $\bm w$. The final score is
	\begin{equation}
		S(\bm w)
		=
		S_0(\bm w)
		\exp[-\gamma\delta_{A^*}(\bm w)],
		\qquad
		\gamma=0.5.
		\label{eq:optimizer_score}
	\end{equation}
	The parameter $\gamma$ controls the strength of the penalty applied when the QUR mean bottleneck SKR falls below the A* reference value. This score function is used in Algorithm~\ref{alg:qur_weights} to pick weight vectors that follow our defined criteria. An example of $S(\bm w)$ for a fixed network configuration is shown in Figure~\ref{fig:montecarlo}.
	
	\subsection*{Machine-learned selection of fixed QUR weights}
	
	QUR$_{\mathrm{go}}$ selects the highest scoring weight vector among
	the tested samples, but requires a new $8,000$ random weight search for every graph. QUR$_{\mathrm{fw}}$ is introduced to avoid this repeated optimization. It uses exactly the same QUR routing
	cost in Eqs.~\ref{eq:qur_transition_cost}--\ref{eq:qur_priority}, but
	replaces the graph specific weights with one weight vector selected from the ensemble of 50 graphs per configuration.
	
	The machine learning model learns how the QUR score changes with the graph
	properties and the six free QUR weights. For graph $g$, we denote the graph
	information by $\bm y_g$ and the $i$th tested weight vector by
	$\bm w_{g,i}$. The input to the model is therefore
	
	\begin{equation}
		\bm z_{g,i}
		=
		\begin{bmatrix}
			\bm y_g\\
			\bm w_{g,i}
		\end{bmatrix},
		\label{eq:ml_input}
	\end{equation}
	
	where $\bm w_{g,i}\in\mathbb R^6$ contains the six free QUR weights.
	
	The vector $\bm y_g$ contains information about the graph on which the
	weights are tested. This includes quantities already defined above, such as
	the number of nodes $N_V$, network radius $R$, degree threshold $\mathcal D$,
	maximum number of paths $M$, and the number of edges. It also contains
	quantities calculated from the particular graph, including the fractions of
	trusted and untrusted nodes, node degree and fiber lengths, the
	source--destination separation, and the corresponding A* results. Thus,
	$\bm y_g$ describes the network conditions under which a QUR weight vector
	is tested.
	
	For every graph, the $8,000$ QUR weight vectors are first evaluated using the
	exact routing model. Each graph--weight pair has an exact score
	$S_{g,i}$ from Eq.~\ref{eq:optimizer_score}. The machine-learning model is
	trained to reproduce how this score changes with $\bm y_g$ and
	$\bm w_{g,i}$. Before training, the score is transformed according to
	
	\begin{equation}
		S'_{g,i}
		=
		\begin{cases}
			\ln(1+10^6), & S_{g,i}=+\infty,\\[3pt]
			\ln(1+S_{g,i}), & 0<S_{g,i}<\infty,\\[3pt]
			0, & \text{otherwise},
		\end{cases}
		\label{eq:ml_target}
	\end{equation}
	
	which reduces the effect of very large score values during training. When all
	50 graphs provide usable QUR results, the model is
	trained using up to 50 $\times$ 8,000 exact graph--weight evaluations.
	
We use histogram based gradient boosting \cite{ml1,ml2}, whose
standard regression tree formulation is summarized in Supplementary Note~5.
The settings are a maximum of $T_{\mathrm{ML}}=300$ trees, a learning rate
$\lambda_{\mathrm{ML}}=0.06$, at most 31 leaf nodes, at least 20 training
samples per leaf, and an $\ell_2$ regularization value of 1.
	
	The trained model does not directly generate the fixed QUR weights. Instead,
	it is used to compare a finite selection of potential weight vectors. For
	each graph, the 500 highest scoring vectors from the $8,000$ QUR trials are
	retained. A further $10,000$
	randomly sampled vectors are added to explore weight
	combinations that may perform well across several graphs even if they were
	not among the highest scoring vectors of any individual graph. The resulting fixed weight vector is chosen from
	
	\begin{equation}
		\mathcal W_\mathrm{fw}
		=
		\bigcup_{g=1}^{G' \leq 50}
		\operatorname{Top}_{500}(\mathcal W_g)
		\;\cup\;
		\mathcal W_{\mathrm{rand}},
		\label{eq:ml_candidate_pool}
	\end{equation}
	
	where $G'$ is the number of usable graphs, $\mathcal W_g$ is the set of $8,000$ weight vectors tested on graph $g$ and $\mathcal W_{\mathrm{rand}}$ contains the additional $10{,}000$ random weight vectors. The model predicts the score of every weight vector in $\mathcal W_{\mathrm{fw}}$ for each graph. These predictions are normalized by the highest predicted score for that graph, and the vector with the highest average normalized score is selected.
	
	\subsection*{Averaging and plotted comparisons}
	
	For each graph $g=1,\ldots,50$, two measures of SKR per node are calculated. The first measure uses
	the mean number of nodes per path,
	
	\begin{equation}
		\eta_{\mathrm{int},g}
		=
		\frac{\overline K_g}{\overline n_g},
	\end{equation}
	
	where $\overline n_g$ is the mean number of nodes per accepted path, which includes repeated nodes, and $\overline K_g$ is the average bottleneck SKR.
	
	The second measure uses the number of unique nodes across all accepted paths,
	
	\begin{equation}
		\eta_{\mathrm{unique},g}
		=
		\frac{K_{\mathrm{sum},g}}
		{n_{\mathrm{unique},g}},
	\end{equation}
	
	where $n_{\mathrm{unique},g}$ is the number of unique nodes used across all accepted paths.
	
	The average SKR-per-node measure is then
	
	\begin{equation}
		\eta_{\mathrm{avg},g}
		=
		\frac{
			\eta_{\mathrm{int},g}
			+
			\eta_{\mathrm{unique},g}
		}{2}.
		\label{eq:average_efficiency_graph}
	\end{equation}
	
	These quantities are calculated separately for each graph before averaging over the graph ensemble. The error bars shown for the curves are the standard errors calculated from the 50 graph values.
	
	Node saving is calculated separately using the mean number of nodes per path and the number of unique nodes,
	
	\begin{align}
		s_{\mathrm{int}}
		&=
		100\left(
		1-
		\frac{
			\langle\overline n\rangle_{\mathrm{QUR}}
		}{
			\langle\overline n\rangle_{\mathrm{ref}}
		}
		\right),
		\nonumber\\
		s_{\mathrm{unique}}
		&=
		100\left(
		1-
		\frac{
			\langle n_{\mathrm{unique}}\rangle_{\mathrm{QUR}}
		}{
			\langle n_{\mathrm{unique}}\rangle_{\mathrm{ref}}
		}
		\right).
	\end{align}
	
	Here, $\langle\cdot\rangle$ denotes the average of the corresponding quantity over the 50 graph ensemble. The subscript ``ref'' denotes the reference routing algorithm, either A* or Max--Min. The average node saving shown in Figure~\ref{fig:overall_qur_comparison} is
	
	\begin{equation}
		s_{\mathrm{avg}}
		=
		\frac{
			s_{\mathrm{int}}
			+
			s_{\mathrm{unique}}
		}{2}.
		\label{eq:average_node_saving}
	\end{equation}
	
	Repeated node use is determined by comparing the total number of nodes used across all accepted paths with the number of unique nodes. For each plotted configuration, the repeated node percentage is
	
	\begin{equation}
		P_{\mathrm{reuse}}
		=
		100 \;\;
		\frac{
			\displaystyle
			\sum_{g=1}^{50}
			\left(
			m_g\overline n_g
			-
			n_{\mathrm{unique},g}
			\right)
		}{
			\displaystyle
			\sum_{g=1}^{50}
			m_g\overline n_g
		},
		\label{eq:reuse_percentage}
	\end{equation}
	
	where $m_g$ is the number of accepted paths in graph $g$. Because accepted paths are internally node disjoint, this quantity measures repeated nodes within individual paths rather than reuse between different accepted paths. Thus, $P_{\mathrm{reuse}}=0$ means that every node used across the accepted paths is unique. A larger value indicates that some nodes are used more than once.

    \subsection*{Use of generative artificial intelligence}
    
    ChatGPT (OpenAI), Gemini (Google), and Claude (Anthropic) were used to assist with code optimization, multiprocessing and caching, development of the machine-learning implementation and figure aesthetics. All AI-assisted outputs were reviewed, tested, and verified by the authors, who retained full responsibility for the final work.
	
	\section*{Data availability}
	
	The data generated during the study is available from the corresponding author upon reasonable request.

	\section*{Code availability}
	
	The code used for this study is available from the corresponding author upon reasonable request.    
	
	\bibliographystyle{unsrtnat}
	\bibliography{references}
	
	\section*{Acknowledgements}
	The authors gratefully acknowledge Adil Umar Khawaja for assistance in developing the Monte Carlo simulation framework used in this work.
	
	\section*{Author contributions}
	F.A. developed the network model, routing framework, simulations and manuscript. A.K. supervised the project and reviewed the manuscript. Both authors discussed the results and approved the final manuscript.
	
	\section*{Competing interests}
	The authors declare no competing interests.
	
\end{document}


\maketitle

This document collects the established theoretical and methodological
background used in the main manuscript. The numerical settings and the
QUR routing and weight selection procedures are specified in the main Methods.

\section*{Supplementary Note~1. Composable finite-size SKR transition model}

We evaluate all physical and relay transitions using the composable finite-size security framework for Gaussian quantum networks with untrusted relays under collective Gaussian attacks \cite{pirandolanetwork}. A transition contains $L_\ell=1,\ldots,4$ physical links, where $L_\ell=1$ denotes a direct P2P connection and $L_\ell>1$ contains $L_\ell-1$ untrusted CV Bell-measurement relays.
	
	For squeezing $r_{\mathrm{dB}}$, the TMSV variance is
	\begin{equation}
		\mu=\cosh(2r), \qquad \mathrm{for} \qquad
		r=\frac{r_{\mathrm{dB}}\ln 10}{20}
	\end{equation}
	and each physical link $\ell$ is represented in standard form by
	\begin{equation}
		\Gamma_\ell=\Gamma(a_\ell,b_\ell,c_\ell),
		\qquad
		a_\ell=\mu,\quad
		b_\ell=T_\ell\mu+1-T_\ell+\xi,\quad
		c_\ell=\sqrt{T_\ell(\mu^2-1)},
	\end{equation}
	where $\xi$ is the output referenced excess noise and transmissivity of the link is
	\begin{equation}
		T_\ell=10^{-\alpha d_\ell/10},
		\qquad
		\alpha=0.2~\mathrm{dB/km}.
	\end{equation}
	where $d_\ell$ is the link distance. For P2P transitions, the SKR is optimized over $r_{\mathrm{dB}}\in[1,30]$ dB. Relay transitions are reduced recursively using the Gaussian swapping map. For a link $\ell = 1$ and $\ell = 2$, the swapped covariance matrix is
	\begin{equation}
		\Gamma(a',b',c') =
		\mathcal S[
		\Gamma(a_1,b_1,c_1),
		\Gamma(a_2,b_2,c_2)],
	\end{equation}
	where
	\begin{align}
		a'&=a_1-\frac{\eta_{\mathrm B}c_1^2}{D},
		\qquad
		b'=b_2-\frac{\eta_{\mathrm B}c_2^2}{D},
		\qquad
		c'=\frac{\eta_{\mathrm B}c_1c_2}{D}, \nonumber \\ 
		D&=\eta_{\mathrm B}(b_1+a_2)+1-\eta_{\mathrm B},
		\label{eq:swap_map}
	\end{align}
	where $\eta_B$ is the efficiency of the Bell measurement device.
	
	For a block of $\mathcal N$ signals,
	\begin{equation}
		m=\lfloor f_{\mathrm{PE}}\mathcal N\rfloor,
		\qquad
		n=\mathcal N-m,
		\qquad
		p_{\mathrm{EC}}=1-\mathrm{FER},
	\end{equation}
	where FER is the frame error rate which gives us $p_\mathrm{EC}$ as the probability that error correction succeeds, and $f_\mathrm{PE}$ is the fraction of signals used for parameter estimation. The parameter estimation uncertainty is
	\begin{equation}
		\delta_{\mathrm{PE}}
		=
		\sqrt{\frac{4\ln(8/\epsilon_{\mathrm{PE}})}{m}}.
	\end{equation}
	The worst case covariance parameters are taken as
	\begin{equation}
		a_{\mathrm{wc}}=a+\delta_{\mathrm{PE}}(a+1),
		\qquad
		b_{\mathrm{wc}}=b+\delta_{\mathrm{PE}}(b+1),
		\qquad
		c_{\mathrm{wc}}
		=c-\frac{\delta_{\mathrm{PE}}}{2}(a+b+2),
	\end{equation}
	The corresponding worst case covariance matrix is
	\begin{equation}
		\Gamma_{\mathrm{wc}}
		=
		\Gamma(a_{\mathrm{wc}},b_{\mathrm{wc}},c_{\mathrm{wc}}).
	\end{equation}
	These worst case covariance matrix parameters give us the symplectic eigenvalues
	\begin{equation}
		\nu_{1,2}
		=
		\sqrt{
			\frac{
				\Delta_{\mathrm{wc}}
				\mp
				\sqrt{\Delta_{\mathrm{wc}}^2-4\det\Gamma_{\mathrm{wc}}}
			}{2}},
	\end{equation}
	where det is the determinant and
	\begin{equation}
		\Delta_{\mathrm{wc}}
		=
		a_{\mathrm{wc}}^2+b_{\mathrm{wc}}^2-2c_{\mathrm{wc}}^2.
	\end{equation}
	For heterodyne detection, the conditional symplectic eigenvalue is
	\begin{equation}
		\nu_{\mathrm{cond}}
		=
		a_{\mathrm{wc}}
		-
		\frac{c_{\mathrm{wc}}^2}{b_{\mathrm{wc}}+1}.
	\end{equation}
	The corresponding worst case mutual and Holevo information, respectively, are
	\begin{align}
		I_{AB}^{\mathrm{wc}}
		&=
		\log_2\!\left(
		\frac{a_{\mathrm{wc}}+1}{\nu_{\mathrm{cond}}+1}
		\right),
		\nonumber\\
		\chi_{BE}^{\mathrm{wc}}
		&=
		g(\nu_1)+g(\nu_2)-g(\nu_{\mathrm{cond}}),
	\end{align}
	where the bosonic entropy function is
	\begin{equation}
		g(\nu)
		=
		\frac{\nu+1}{2}\log_2\!\left(\frac{\nu+1}{2}\right)
		-
		\frac{\nu-1}{2}\log_2\!\left(\frac{\nu-1}{2}\right).
	\end{equation}
	The key rate after incorporating parameter estimation uncertainty is
	\begin{equation}
		K_{\mathrm{PE}}
		=
		\beta I_{AB}^{\mathrm{wc}}
		-
		\chi_{BE}^{\mathrm{wc}}.
	\end{equation}
	The composable finite size SKR assigned to the transition $\ell$ is
	\begin{equation}
		K_\ell
		=
		\frac{np_{\mathrm{EC}}}{\mathcal N}
		\left[
		K_{\mathrm{PE}}
		-
		\frac{\Delta_{\mathrm{AEP}}}{\sqrt n}
		+
		\frac{\Theta}{n}
		\right],
		\label{eq:composable_rate}
	\end{equation}
	where the finite sized correction terms are
	\begin{align}
		\Delta_{\mathrm{AEP}}
		&=
		4\log_2(\sqrt{|\mathcal X|}+2)
		\sqrt{\log_2(18p_{\mathrm{EC}}^{-2}\epsilon_s^{-4})},
		\nonumber\\
		\Theta
		&=
		\log_2 \left[
		p_{\mathrm{EC}}
		\left(1-\frac{\epsilon_s^2}{3}\right)
		\right]
		+
		2\log_2(\sqrt2 \epsilon_h),
	\end{align}
	where $\epsilon_c$, $\epsilon_s$, $\epsilon_h$ and $\epsilon_{\mathrm{PE}}$ are the security failure probabilities associated with correctness, smoothing, hashing and parameter estimation respectively and $|\mathcal X|$ is the size of the discretized measurement alphabet. The corresponding security parameter for the transition is
	\begin{equation}
		\epsilon_{\mathrm{tr}}
		=
		\epsilon_c+\epsilon_s+\epsilon_h
		+p_{\mathrm{EC}}\epsilon_{\mathrm{PE}}.
	\end{equation}

The numerical parameter used in this study are specified in the main Methods.

\section*{Supplementary Note~2. Waxman graph generation}

The standard Waxman construction used in the simulations is summarized below \cite{harney}.

The physical network is modeled as an undirected Waxman graph $G=(V,E)$ with $V$ vertices and $E$ edges, and $N_V$ nodes are uniformly distributed within a disk of radius $R$. For each node $i$, the radial and angular coordinates are
	generated as
	\begin{equation}
		\rho_i=R\sqrt{U_i},
		\qquad
		\theta_i=2\pi V_i,
		\qquad
		\bm x_i=\rho_i(\cos\theta_i,\sin\theta_i),
	\end{equation}
	where $U_i,V_i\sim\mathcal U(0,1)$ and $\bm x_i$ is the resulting
	position of node $i$. Edges are generated according to
	\begin{equation}
		P[(u,v)\in E]
		=
		\beta_W
		\exp \left(-\frac{d_{uv}}{r_0}\right),
		\qquad
		d_{uv}=\|\bm x_u-\bm x_v\|_2,
		\label{eq:waxman}
	\end{equation}
	where $u$ and $v$ are any two nodes in the network, $d_{uv}$ is their Euclidean distance, and $\|\cdot\|_2$ denotes the Euclidean norm. The parameter $r_0=0.07R$ is the characteristic connection distance and controls how rapidly the probability of an edge decreases with node separation, while $\beta_W$ controls the overall connection probability of the Waxman graph. Each edge represents an optical fiber link with attenuation coefficient $\alpha=\SI{0.2}{dB\per km}$.

\section*{Supplementary Note~3. A* routing}

The geometric benchmark uses A* search \cite{astar}. Route notation follows from the main Methods.

A* is the geometric shortest route benchmark. The cost of a transition
	$\ell$ is the total length $d_{\ell}$ of the fiber links contained
	within that transition. For a partial route $\mathcal R_{S\rightarrow v}$
	ending at trusted node $v$,
	\begin{equation}
		C_{A^*}(v)
		=
		\sum_{\ell\in\mathcal R_{S\rightarrow u}}d_{\ell} + d_{\ell}(u,v).
	\end{equation}
	The search priority is
	\begin{equation}
		f_{A^*}(v)
		=
		C_{A^*}(v)
		+
		\|\bm x_v-\bm x_T\|_2,
	\end{equation}
	where the second term is the Euclidean distance from current trusted node $v$ to the destination $T$. The next trusted node is selected according to the lowest value of $f_{A^*}$.

\section*{Supplementary Note~4. Max--Min routing}

The bottleneck rate benchmark uses Max--Min routing \cite{maxmin}. Here, $K(\mathcal R)=\min_{\ell\in\mathcal R}K_\ell$, with route notation as in the main Methods.

Max--Min selects the route with the largest bottleneck SKR,
	\begin{equation}
		\mathcal R_{\mathrm{MM}}^*
		=
		\arg\max_{\mathcal R\in\mathcal R_{ST}}
		K(\mathcal R),
		\label{eq:maxmin}
	\end{equation}
	where $\mathcal R_{ST}$ is the set of available routes between the source and destination. If the current node is $u$ and $v$ is a reachable node, extending a partial route $\mathcal R_{S\rightarrow u}$ through the transition $\ell=(u,v)$ gives
	\begin{equation}
		K(\mathcal R_{S\rightarrow v})
		=
		\min\!\left[
		K(\mathcal R_{S\rightarrow u}),
		K_\ell
		\right].
	\end{equation}
	
	For each reachable node $v$, the partial route with the largest bottleneck SKR is retained. The Max--Min search priority is therefore
	\begin{equation}
		f_{\mathrm{MM}}(v)
		=
		\max_{\mathcal R\in\mathcal R_{S\rightarrow v}}
		K(\mathcal R),
	\end{equation}
	and the reachable node with the largest $f_{\mathrm{MM}}(v)$ is explored next.

\section*{Supplementary Note~5. Gradient boosting score model}

Here, $\bm z$ combines graph properties and a QUR weight vector, and $S'$ denotes the transformed optimization score defined in the main Methods.

We use histogram based gradient boosting \cite{ml1,ml2}. The model is formed
	from a sequence of regression trees. Each new tree provides a correction to
	the prediction obtained from the preceding trees. Since the model is trained
	using the transformed score $S'$, its final prediction is written as
	
	\begin{equation}
		S^{\prime,\mathrm{ML}}(\bm z)
		=
		F_0
		+
		\lambda_{\mathrm{ML}}
		\sum_{t=1}^{T_{\mathrm{ML}}}
		f_t(\bm z),
		\label{eq:boosted_model}
	\end{equation}
	
	where $F_0$ is the initial prediction, $f_t(\bm z)$ is the correction
	provided by regression tree $t$, $T_{\mathrm{ML}}=300$ is the maximum
	number of trees, and $\lambda_{\mathrm{ML}}=0.06$ controls
	the contribution of each new tree. The model parameters are chosen to limit
	the complexity of the regression trees and reduce overfitting, with a maximum
	of 31 leaf nodes, at least 20 training samples per leaf, and an $\ell_2$
	regularization value of 1.

\bibliographystyle{unsrtnat}
\bibliography{references}